\documentclass[12pt,epsfig]{article}
\usepackage{amsmath, amssymb}
\usepackage[dvips]{graphicx,color}

\def\lromn#1{\uppercase\expandafter{\romannumeral#1}}

\newcommand{\del}{\partial}

\begin{document}
\begin{titlepage}
\begin{center}


\hfill  OU-HET 640/2009 \\
\hfill UT-HET 031\\

\vspace{0.5cm}
{\LARGE\bf Gauge-Higgs Dark Matter}

\vspace{1cm}
{\bf Naoyuki Haba}$^{(a),}$
\footnote{E-mail: haba@het.phys.sci.osaka-u.ac.jp},
{\bf Shigeki Matsumoto}$^{(b),}$
\footnote{E-mail: smatsu@sci.u-toyama.ac.jp},
{\bf Nobuchika Okada}$^{(c),}$
\footnote{E-mail: okadan@ua.edu}, \\
and
{\bf Toshifumi Yamashita}$^{(d),}$
\footnote{E-mail: yamasita@eken.phys.nagoya-u.ac.jp}

\vspace{1cm}
{\it
$^{(a)}${Department of Physics, Osaka University,
        Toyonaka, Osaka 560-0043, Japan} \\
$^{(b)}${Department of Physics, University of Toyama, Toyama 930-8555, Japan} \\
$^{(c)}${Department of Physics and Astronomy, 
University of Alabama, \\
Tuscaloosa, AL 35487, USA
} \\
$^{(d)}${Department of Physics, Nagoya University, Nagoya 464-8602, Japan}
}

\vspace{1cm}
{\bf Abstract}
\end{center}

When the anti-periodic boundary condition is imposed 
 for a bulk field in extradimensional theories, 
 independently of the background metric, 
 the lightest component in the anti-periodic field 
 becomes stable and hence a good candidate 
 for the dark matter in the effective 4D theory 
 due to the remaining accidental discrete symmetry. 
Noting that in the gauge-Higgs unification scenario, 
 introduction of anti-periodic fermions 
 is well-motivated by a phenomenological reason, 
 we investigate dark matter physics in the scenario. 
As an example, we consider a five-dimensional 
 SO(5)$\times$U(1)$_X$ gauge-Higgs unification model 
 compactified on the $S^1/Z_2$ with the warped metric. 
Due to the structure of the gauge-Higgs unification, 
 interactions between the dark matter particle and 
 the Standard Model particles are largely 
 controlled by the gauge symmetry, and hence 
 the model has a strong predictive power 
 for the dark matter physics.  
Evaluating the dark matter relic abundance, 
 we identify a parameter region consistent 
 with the current observations. 
Furthermore, we calculate the elastic scattering cross section 
 between the dark matter particle and nucleon 
 and find that a part of the parameter region 
 is already excluded by the current experimental results 
 for the direct dark matter search and most of 
 the region will be explored in future experiments. 


\end{titlepage}
\setcounter{footnote}{0}

\section{Introduction}

The standard model (SM) of particle physics successfully 
 explains almost of all the experimental results 
 around the electroweak scale.
Nevertheless, the SM suffers from several problems and 
 this fact strongly motivates us to explore physics beyond the SM.
One of them is the so-called hierarchy problem 
 originating from the ultraviolet sensitivity 
 of the SM Higgs doublet mass, and another one 
 is the absence of candidates for the dark matter particle. 
In this paper we propose an extra-dimensional scenario 
 which can provide a possible solution to these two problems.

Among many models proposed to solve the hierarchy problem, 
 we concentrate on the gauge-Higgs unification 
 scenario~\cite{Manton:1979kb,YH}.
In this scenario, the SM Higgs doublet field is identified 
 with an extra-dimensional component of the gauge field 
 in higher-dimensional gauge theories 
 where the extra spacial dimensions are compactified to 
 realize four-dimensional effective theory at low energies. 
The higher-dimensional gauge symmetry protects the Higgs 
 doublet mass from ultraviolet divergences~\cite{YH,finiteness}, 
 and hence the hierarchy problem can be solved. 
In the context of the gauge-Higgs unification scenario, 
 many models have been considered 
 in both the flat~\cite{Csaki}-\cite{GGHU} and 
 the warped~\cite{RS} background geometries~\cite{GHUinRS}-\cite{pNG}.
However, the latter problem has not been investigated in this scenario, 
 except for a few literatures~\cite{DMinGHU,Carena:2009yt,hosotani},
 and in this paper, we propose a dark matter candidate 
 which can be naturally incorporated 
 in the gauge-Higgs unification scenario.

In the next section, we show a simple way to introduce 
 a candidate for the dark matter particle 
 in general higher-dimensional models. 
In a sharp contrast with the usual Kaluza-Klein (KK) dark matter
 in the universal extradimension scenario~\cite{KKDM}, 
 our procedure is independent of the background 
 space-time metric. 
In section 3, we apply this to the gauge-Higgs unification 
 scenario and show that a dark matter candidate as a 
 weakly-interacting-massive-particle (WIMP) emerges. 
For our explicit analysis, we consider a gauge-Higgs unification model 
 based on the gauge group SO(5)$\times$U(1)$_X$ 
 in five-dimensional warped background metric 
 with the fifth dimension compactified on the $S^1/Z_2$ orbifold. 
In section 4, we evaluate the relic abundance 
 of the dark matter particle and its detection rates 
 in the direct dark matter detection experiments. 
Section 5 is devoted to summary.

\section{A new candidate for the dark matter}
\label{Sec:APDM}

A stable and electric charge neutral WIMP is 
 a suitable candidate for the dark matter. 
In general, a certain symmetry (parity) is necessary 
 to ensure the stability of a dark matter particle. 
Such a symmetry can be imposed by hand in some models 
 or it can be accidentally realized  
 such as the KK parity~\cite{KKDM}. 
The KK parity is actually an interesting possibility 
 for introducing a dark matter candidate 
 in higher-dimensional models. 
However, we need to elaborate a model in order to 
 realize the KK parity in general warped 
 background geometry~\cite{Agashe:2007jb}. 
In a simple setup, the KK parity is explicitly broken 
 by a warped background metric and 
 the KK dark matter is no longer stable~\cite{OY}. 
So, here is an interesting question: 
Is it possible in extradimensional models to introduce 
 a stable particle independently of the background
 space-time metric, without imposing any symmetries by hand? 
In the following we address our positive answer to this question. 
In fact, when we impose the anti-periodic (AP) boundary condition 
 on bulk fields, the lightest AP field turns out to be stable.

In models with the toroidal compactification, no matter what 
 further orbifoldings are, the Lagrangian ${\mathcal L}$ 
 should be invariant under a discrete shift of the coordinate 
 of the compactified direction, 
\begin{equation}
 {\cal L}(x,y+2\pi R) = {\cal L}(x,y),
\end{equation}
 where $x$ and $y$ denote the non-compact four dimensional coordinate 
 and the compact fifth-dimensional one with a radius $R$, respectively. 
When we introduce some fields which have the AP boundary condition as 
\begin{equation}
 \Phi(x,y+2\pi R) = -\Phi(x,y),
\end{equation}
 these fields never appear alone but always do in pairs in the Lagrangian,
 since the Lagrangian must be periodic.
Thus, there exists an accidental $Z_2$ parity, under which
 the AP (periodic) fields transform as odd (even) fields.
This concludes that the lightest AP field is stable\footnote{
Similarly to the KK parity, 
 Lagrangian on the boundaries must be restricted 
 to respect the $Z_2$ parity. 
} 
 and can be a good candidate for the dark matter 
 if it is colorless and electric-charge neutral.

In this way, a dark matter candidate can be generally incorporated 
 as the lightest AP field in higher-dimensional models.
However, except for providing the dark matter candidate, 
 there may be no strong motivation for introducing such AP fields. 
In fact, AP fields often plays a crucial role 
 in the gauge-Higgs unification scenario 
 to make a model phenomenologically viable, 
 and therefore a dark matter candidate is simultaneously 
 introduced in such a model.

\section{Gauge-Higgs Dark Matter}
\label{Sec:GHDM}

We show a model of the gauge-Higgs unification, 
 which naturally has a dark matter candidate. 
The dark matter particle originates from an AP field which is introduced 
 in a model for a phenomenological reason as will be discussed below. 

We know well that, it is difficult, 
 in simple gauge-Higgs unification models with the flat metric, 
 to give a realistic top quark mass and a Higgs boson mass 
 above the current experimental lower bound. 
This difficulty originates from the fact that 
 effective Higgs potential in the gauge-Higgs unification model 
 results in the Wilson line phase of order one. 
%
%
When we consider the gauge-Higgs unification scenario 
 in the warped metric of the extra dimension, 
 this problem can be solved because  of the effect of 
 the warped metric, although the Wilson line phase 
 of order one is obtained from effective Higgs potential. 
However, as is claimed in Ref.~\cite{EWPMinGHU}, 
 a small Wilson line phase is again required 
 in order for the scenario to be consistent with 
 the electroweak precision measurements. 
Therefore, it is an important issue 
 in the gauge-Higgs unification scenario 
 how to naturally obtain a small Wilson line phase.

A simple way is to introduce AP fermions in a model. 
It has been shown in Ref.~\cite{APinGHU} that 
 a small Wilson line phase is actually obtained 
 by introducing AP fermions. 
This is the motivation we mentioned above\footnote{
In Ref.~\cite{DMinGHU},
 with a {\it similar} purpose, a {\it similar} $Z_2$ symmetry
 is imposed {\it but}
 by hand.}.
An AP fermion, once introduced, not only reduces 
 unwanted new particle effects to the precisely measured 
 SM parameters but also provides a dark matter candidate 
 as its lightest electric-charge neutral component. 
We call the dark matter candidate in the AP fermion 
 ``gauge-Higgs dark matter'' in this paper. 
The interactions between the dark matter and the Higgs field is 
 largely controlled by the gauge symmetry,  
 since the Higgs field is a part of the gauge field 
 in the gauge-Higgs unification scenario. 
This fact leads to a strong predictive power of the model 
 for the dark matter phenomenology.

\subsection{A model}
\label{Sec:model}

Here we explicitly examine a 5D gauge-Higgs unification model
 with a dark matter particle. 
The model is based on the gauge symmetry 
 SO(5)$\times$U(1)$_X$~\cite{Carena:2009yt,hosotani}
 compactified on the simplest orbifold $S^1/Z_2$ 
 with the warped metric~\cite{RS} 
\begin{eqnarray}
 {\rm d} s^2
 =
 G_{M N} {\rm d} x^M {\rm d} x^N
 =
 e^{-2\sigma(y)} \eta_{\mu \nu} {\rm d} x^\mu {\rm d} x^\nu
 -
 {\rm d} y^2,
\end{eqnarray}
where $M=0,1,2,3,5$, $\mu=0,1,2,3$,
 $\sigma(y) = k |y|$ at $-\pi R \leq y \leq \pi R$, 
 $\sigma(y) = \sigma(y + 2 \pi R)$, and  
 $\eta_{\mu \nu} = {\rm diag}(1, -1, -1, -1)$ is 
 the 4D flat metric. 
We define the warp factor $a= \exp(-\pi k R)$
 and as a reference value, we set the curvature $k$ and 
 the radius $R$ to give the warp factor  $a  = 10^{-15}$.

The bulk SO(5) gauge symmetry is broken down to
 SO(4)$\simeq$SU(2)$_L\times$SU(2)$_R$
 by the boundary conditions~\cite{Kawamura}.
Concretely, the gauge field and its 5th component transform
 around the two fixed points $y_0=0$ and $y_L=\pi R$ as
\begin{eqnarray}
 A_\mu(x,\,y_i-y) &=& P_i A_\mu(x,\,y_i+y) P_i^\dagger, \\
 A_5(x,\,y_i-y) &=& -P_i A_5(x,\,y_i+y) P_i^\dagger,
\end{eqnarray}
 under the $Z_2$ parity, 
 where 
 $P_0=P_L={\rm diag.}(-1,-1,-1,-1,+1)$ for the five-by-five anti-symmetric
 matrix representation of the generators acting on the vector
 representation, $\bf5$.
As for the remaining SO(4)$\times$U(1)$_X$ gauge symmetry,
 the SU(2)$_R\times$U(1)$_X$ is assumed to be broken down to the hypercharge
 symmetry U(1)$_Y$
 by a VEV of an elementary Higgs field\footnote{
Note that introducing the elementary Higgs field 
 at the $y=0$ orbifold fixed point has no contradiction 
 against the motivation of the gauge-Higgs unification scenario 
 since the mass of the Higgs fields and their VEVs are of 
 the order of the  Planck scale.
In this case, they decouple from TeV scale physics. 
} 
%
put on the $y=0$ orbifold fixed point.
Now the remaining gauge symmetry is the same as the SM, 
 where there exists the zero-mode of $A_5$ 
 which is identified as the SM Higgs doublet
 (possessing the right quantum numbers). 
When the zero mode of $A_5$ develops a non-trivial VEV,
 the SO(4) symmetry is broken down to SO(3)$\simeq$SU(2)$_D$ which
 is the diagonal part of SU(2)$_L\times$SU(2)$_R\simeq$SO(4).
Taking the boundary Higgs VEV into account,
 the electromagnetic U(1)$_{\rm EM}$ is left with unbroken. 
Thanks to the custodial symmetry which is violated 
 only at the $y=0$ fixed point, that is, a superheavy energy scale,
 the correction to the $\rho$-parameter is naturally
 suppressed~\cite{EWPMinGHU}. 
This allows the KK scale as low as a few TeV 
 without any contradictions against current experiments.

The components of gauge field are explicitly written as
\begin{equation}
 A_M=
 \left(
 \begin{array}{cccc|c}
 0 & A_V^3 & -A_V^2 & A_A^1 & A_H^1 \\
  &   0   &  A_V^1 & A_A^2 & A_H^2 \\
  &       &    0   & A_A^3 & A_H^3 \\
  &       &        &   0   & A_H^4 \\ \hline
  &       &        &       &   0
 \end{array}
 \right)_M,
\end{equation}
 where
\begin{eqnarray}
 A_{\scriptsize\begin{array}{c}V\vspace{-2mm}\\A\end{array}}^i =
 \frac1{\sqrt2}(A_L^i\pm A_R^i),\qquad (i=1,\,2,\,3), \\
 A_F^\pm = \frac1{\sqrt2}(A_F^1\mp i A_F^2),\qquad (F=V,A,H). 
\end{eqnarray}
The zero-modes of $A_5$ exist on $A_H$ and its VEV 
 can be rotated into only $(A_H^4)_5$ component 
 by the SO(4) symmetry, by which 
 the Wilson line phase $\theta_W$ is defined as
\begin{eqnarray}
 W \equiv e^{i \theta_W}
 =
 P \exp\left( {-i g \int^{\pi R}_{-\pi R} {\rm d} y ~G^{55} (A_H^4)_5} \right),
 \label{WilsonLinePase}
\end{eqnarray}
where $P$ denotes the path ordered integral.
For vanishing $\theta_W$,
 the SM gauge bosons are included 
 in $A_L$ and $B_X$ (which is 
 the gauge boson of the U(1)$_X$ symmetry), 
 while the $A_H$ component is mixed into 
 the mass eigenstates of weak bosons for non-vanishing $\theta_W$.

We do not specify the fermion sector of the model 
 but just assume it works well, since this sector 
 is not strongly limited by the gauge symmetry
 and has a lot of model-dependent degrees of freedom.
Thus, in our following analysis we leave the Higgs boson mass $m_h$ and 
 the Wilson line phase $\theta_W$ as free parameters, 
 which should be calculated through the loop induced effective
 potential~\cite{EffPot}-\cite{EffPotRS} 
 once the fermion sector of the model is completely fixed. 
%
%

Let us now consider an AP fermion, $\psi$, 
 as a ${\bf5}_0$-multiplet under SO(5)$\times$U(1)$_X$, 
 in which the dark matter particle is contained. 
A parity odd bulk mass parameter $c$ of this multiplet 
 is involved as an additional parameter~\cite{GherghettaPomarol}. 
The wave function profile along the compactified direction 
 is written by the Bessel functions with the index
 $\alpha=\left|\gamma_5 c+1/2\right|$~\cite{GherghettaPomarol} 
 and the localization of the bulk fermion is controlled 
 by the bulk mass parameter. 
We choose the boundary conditions of this multiplet
 so that the singlet component
 of the SO(4) is lighter than the vector one for small $\theta_W$
 with $ c > 0$.

After the electroweak symmetry breaking, 
 the forth and fifth components are mixed with each other 
 through the non-vanishing Wilson line phase in $(4,5)$ component, 
 while the first, second and third ones are not. 
The combinations of forth and fifth components make up  
 two mass eigenstates: 
The lighter one is nothing but the dark matter particle, $\psi_{\rm DM}$, 
 and we denote the heavier state as $\psi_S$. 
The first, second and third components denoted 
 as $\psi_i$\,$(i=1,\,2,\,3)$ have nothing to do 
 with the electroweak symmetry breaking, and thus 
 degenerate up to small radiative corrections. 
They are heavier than $\psi_S$. 
Note that only dark matter particles themselves have no couplings 
 with the weak gauge bosons, 
 which are linear combinations of $A_V$, $A_A$ and $A_H$ (and $B_X$), 
 and the couplings between the dark matter particle and the weak gauge bosons  
 are always associated  with the transition from/to the heavier partners, 
 $\psi_i$. 
On the other hand,
 both types of couplings exist among 
 the dark matter particle and the Higgs boson. 
At the energy scale below the 1st KK mode mass, 
 the effective Lagrangian is expressed as
\begin{eqnarray}
 {\cal L}^{\rm 4D}_{\rm DM}&=&
  \sum_{i=1,2,3,S,{\rm DM}}
    \bar\psi_i (i\del\hspace{-2.3mm}/ -m_a)\psi_i
 +y_{\rm DM} \bar\psi_{\rm DM} H \psi_{\rm DM}
 \nonumber\\
&& +\bar\psi_S H
     \left(y_S+y_P\gamma_5\right)
    \psi_{\rm DM}
 +\bar\psi_{\rm DM} H
     \left(y_S-y_P\gamma_5\right)
    \psi_S
 \nonumber\\
&& +\sum_{i=1,2,3}\bar\psi_i
    W_i\hspace{-4mm}/\,\,\,
     \left(g^V_i+g^A_i\gamma_5\right)
    \psi_{\rm DM}
 +\bar\psi_{\rm DM}
    W_i\hspace{-4mm}/\,\,
     \left(g^V_i+g^A_i\gamma_5\right)
    \psi_i,
\label{effL}
\end{eqnarray}
 where we denote $Z$ as $W^3$, and set $g_1^h=g_2^h$ due to the remaining
 U(1)$_{\rm EM}$ symmetry.

Once we fix the free parameters $\theta_W$ and $c$ (also the warp factor), 
 we can solve the bulk equations of motion for $A_M$ and $\psi$ 
 (see for example Ref.~\cite{sakamura-unitariy}) 
 and obtain the mass spectra of all the states 
 and effective couplings in Eq.(\ref{effL}) among AP fields, 
 the gauge bosons and the Higgs boson, 
 independently of the Higgs boson mass $m_h$ 
 (which is another free parameter of the model as mentioned above). 
Using calculated spectra and the effective couplings,
 we investigate phenomenology of the gauge-Higgs dark matter 
 in the next section. 
Since we have only three parameters 
 (or four if we count also the warp factor), 
 the model has a strong predictive power. 
%
%

\subsection{Constraints}
\label{Sec:constraints}

Before investigating the gauge-Higgs dark matter phenomenology, 
 we examine an 
 experimental constraint on the Wilson line phase $\theta_W$~\footnote{
Constraints in the case with the flat metric is discussed 
in Ref.~\cite{ConstraintInFlatGHU}. 
}.
In Ref.~\cite{EWPMinGHU}, 
 it is claimed that $\theta_W$ should be smaller than $0.3$ 
 or the KK gauge boson mass larger than 3 TeV 
 in order to be consistent with the electroweak precision measurements. 
Using the relation between $m_W$ and $m_{KK}\equiv\pi k a$ 
 (see for example Ref.~\cite{sakamura-unitariy}), 
\begin{equation}
 m_W \simeq \frac{\theta_W}{\sqrt{\ln(a^{-1})}}\frac{m_{KK}}\pi,
\end{equation}
 and the formula for the first KK gauge boson mass $m_1=0.78m_{KK}$, 
 the latter constraint is translated as $\theta_W\lesssim0.4$. 
%

According to these bounds, we restrict our analysis 
 in the range of a small Wilson line phase, 
 namely, $\theta_W\leq\pi/10$. 
We expect that AP fields not only provide the dark matter particle 
 but also is helpful to realize such small value of $\theta_W$. 

\section{Phenomenology of gauge-Higgs dark matter}

Now we are in a position to investigate 
 the gauge-Higgs dark matter phenomenology. 
We first estimate the relic abundance of the dark matter 
 and identify the allowed region of the model parameter space 
 which predicts the dark matter relic density consistent with 
 the current cosmological observations. 
Furthermore, we calculate the cross section of the elastic 
 scattering between the dark matter particle and nucleon 
 to show implications of the gauge-Higgs dark matter scenario 
 for the current and future direct dark matter detection experiments.

\subsection{Relic abundance}

In the early universe, the gauge-Higgs dark matter is 
 in thermal equilibrium through the interactions 
 with the SM particles. 
According to the expansion of the universe, 
 temperature of the universe goes down and the dark matter eventually 
 decouples from thermal plasma of the SM particles 
 in its non-relativistic regime. 
The thermal relic abundance of the dark matter can be evaluated 
 by solving the Boltzmann equation,
\begin{eqnarray} 
 \frac{d Y}{dx} = 
 -\frac{s \langle \sigma v \rangle}{x H} 
  \left( 1- \frac{x}{3 } \frac{d \log g_{*s}}{d x } \right) 
  \left( Y^2 -Y_{EQ}^2  \right), 
\end{eqnarray}  
where $x=m_{\rm DM}/T$, 
 $\langle \sigma v \rangle$ 
 is the thermal averaged product of  
 the dark matter annihilation cross section ($\sigma $) 
 and the relative velocity of annihilating dark matter particles ($v$), 
 $Y(\equiv n/s)$ is the yield defined as the ratio 
 of the dark matter number density $(n)$ to the entropy density 
 of the universe $(s)$, and the Hubble parameter $H$ 
 is described as $H= \sqrt{(8 \pi/3) G_N \rho}$ 
 with the Newton's gravitational constant 
 $G_N=6.708 \times 10^{39}$ GeV$^{-2}$ 
 and the energy density of the universe ($\rho$). 
The explicit formulas for the number density of the dark matter particle, 
 the energy density, and the entropy density are given, 
 in the Maxwell-Boltzmann approximation, by 
\begin{equation} 
 n = \frac{g_{DM}}{2 \pi^2} \frac{K_2(x)}{x} m^3, \quad 
 \rho = \frac{\pi^2}{30} g_* T^4, \quad 
 s = \frac{2 \pi^2}{45} g_{*s} T^3, 
\end{equation} 
where $K_2$ is the modified Bessel function of the second kind, 
 $g_{DM}=4$ is the spin degrees of freedom 
 for the gauge-Higgs dark matter, and 
 $g_*$ ($g_{*s}$) is the effective massless degrees of freedom 
 in the energy (entropy) density, respectively.

In non-relativistic limit, the annihilation cross section 
 can be expanded with respect to a small relative velocity as 
\begin{eqnarray} 
 \sigma v = \sigma_0 + \frac{1}{4} \sigma_1 v^2 +\cdots,
\end{eqnarray} 
where $v \simeq 2 \sqrt{1-4 m^2/s}$ in the center-of-mass frame 
 of annihilating dark matter particles. 
The first term corresponds to the dark matter annihilations 
 via $S$-wave, while the second is contributed by the $S$- 
 and $P$-wave processes. 
In the Maxwell-Boltzmann approximation, 
 the thermal average of the annihilation cross section 
 is evaluated as 
\begin{eqnarray} 
 \left< \sigma v \right>
 &\equiv&\frac1{8x^4 K_2(x)^2}
   \int_{4x^2}^\infty ds\,
       \sqrt{s}(s-4x^2)K_1(\sqrt{s})\sigma_{ann} \\
 &=& \sigma_0 +\frac32\sigma_1 x^{-1}+\cdots,
\label{left< v}
\end{eqnarray}
where a unit $T=1$ is used in the first line. 
There are several dark matter annihilation modes 
 in both $S$-wave and $P$-wave processes 
 (see Eq.~(\ref{effL})), 
 such as 
 $ \bar{\psi}_{\rm DM} \psi_{\rm DM} \to W^+ W^-, ZZ, HH$ 
 through $\psi_i$, $\psi_S$ and $\psi_{\rm DM}$ exchanges in the $t$-channel 
 and $ \bar{\psi}_{\rm DM} \psi_{\rm DM} \to f \bar{f}, W^+ W^-, ZZ, HH$ 
 through the Higgs boson exchange in the $s$-channel, 
 where $f$ stands for quarks and leptons.  
Once the model parameters, $\theta_W$, $c$ and $m_h$, are fixed, 
 magnitudes of $\sigma_0$ and $\sigma_1$ are calculated.

With a given annihilation cross section, 
 the Boltzmann equation can be numerically solved. 
The relic density of the dark matter is obtained as 
 $\Omega_{\rm DM} h^2 = m_{\rm DM} s_0 Y(\infty)/(\rho_c/h^2)$ 
 with $s_0=2889$ cm$^{-3}$ and 
 $\rho_c/h^2=1.054\times 10^{-5}$ GeV cm$^{-3}$. 
Here, we use an approximate formula~\cite{DMabundanceApp} 
 for the solution of the Boltzmann equation:
\begin{equation}
 \Omega_{\rm DM}h^2 =
 8.766\times10^{-11}({\rm GeV}^{-2})
 \left(\frac{T_0}{2.75 {\rm K}}\right)^3
 \frac{x_f}{\sqrt{g_{*}(T_f)}} 
 \left(\frac12\sigma_0+\frac38\sigma_1 x_f^{-1}\right)^{-1},
\label{Omegah2App}
\end{equation}
 where $x_f=m_{\rm DM}/T_f$ is the freeze-out temperature 
 normalized by the dark matter mass, and $T_0=2.725$ K is the present 
 temperature of the universe. 
The freeze-out temperature is approximately determined by~\cite{DMabundanceApp}
\begin{eqnarray}
 \sqrt{\frac\pi{45G_N}}\frac{45g_{\rm DM}}{8\pi^4}
 \frac{\pi^{1/2}e^{-x_f}}{g_{\rm *s}(T_f)x_f^{1/2}}
 g_{\rm *}(T_f) 
 m_{\rm DM}
 \left(\frac12\sigma_0+\frac34\sigma_1 x_f^{-1}\right)
 \delta(\delta+2)=1.
\label{xfApp}
\end{eqnarray} 
Here the parameter $\delta$ defines $T_f$ through a relation 
 between the yield $Y$ and its value in thermal equilibrium, 
 $Y - Y_{\rm EQ} = \delta Y_{\rm EQ}$, 
 whose value is chosen so as to keep this approximation good. 
We set $\delta=1.5$ according to Ref.~\cite{DMabundanceApp}. 
In these approximations, we include the factor $1/2$ 
 due to the Dirac nature of the gauge-Higgs dark matter 
 (see the discussion below Eq.~(2.16) of Ref.~\cite{DMabundanceApp}).

Let us now compare the resultant dark matter relic density  
 for various $ \theta_W $ and $c$ 
 with the observed value~\cite{DMabundance}:
\begin{equation}
  \Omega_{\rm DM}h^2=0.1143\pm 0.0034.
\end{equation}
\begin{figure}[t]
\begin{center}
\vspace{4mm}
\hspace{-8mm}
\includegraphics[width=0.58\textwidth]{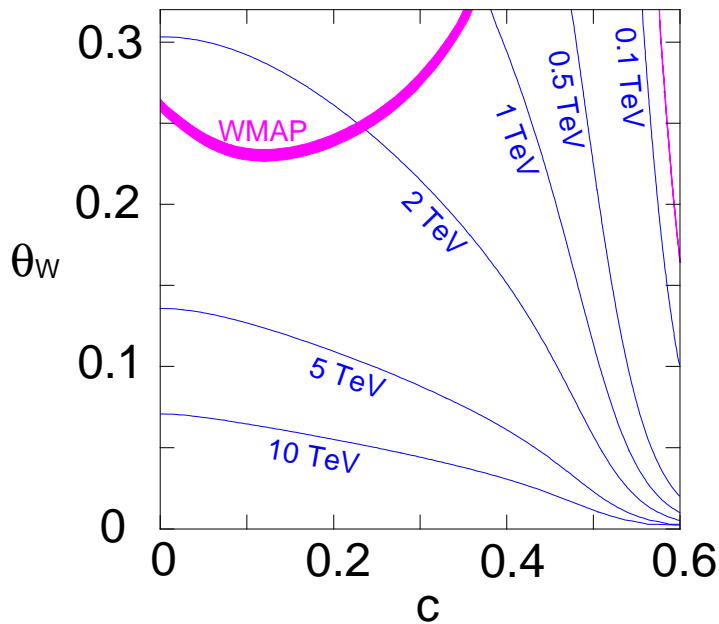}
\end{center}
\caption{{\it The relic abundance}:
 The relic abundances consistent with the observations are obtained in the two red regions. 
 In the upper-left corner outside the red region, the relic abundance is predicted to be too little, 
  while over-abundance of the dark matter relic density is obtained in the other region. 
 The contours corresponding to fixed dark matter masses are also shown. 
 Here, the Higgs mass has been taken to be $m_h=120$ GeV.
}
\label{fig:abundance}
\end{figure}
The result is depicted in Figure \ref{fig:abundance}, 
 where the Higgs boson mass is set as $m_h=120$ GeV. 
The regions consistent with the observations are indicated by red, 
 while too little (much) abundances 
 are obtained in the region above the red line on the upper-left 
 (in the other regions). 
There are two allowed regions: 
One is the very narrow region in upper-right, 
 where the right relic abundance is achieved 
 by the enhancement of the annihilation cross section 
 through the $s$-channel Higgs boson resonance, 
 so that the dark matter mass is $m_{\rm DM} \simeq m_h/2 =60$ GeV there. 
The other one appears in upper-left with 
 the dark matter mass around a few TeV, where dark matter particles 
 can efficiently annihilate into the weak gauge bosons
 and the Higgs bosons through the processes 
 with heavy fermions in the $t$-channel.

\subsection{Direct detection}

Next we investigate the implication of the gauge-Higgs dark matter 
 for the direct detection experiments~\cite{TrAnom}. 
A variety of experiments are underway to directly detect
 dark matter particles through their elastic scatterings off nuclei.
The most stringent limits on the (spin-independent) elastic
 scattering cross section have been reported by the recent 
 XENON10~\cite{XENON10} and CDMS II~\cite{CDMS} experiments: 
 $\sigma_{el}({\rm cm}^2) \lesssim 
 7 \times 10^{-44} - 5 \times 10^{-43}$, 
 for a dark matter mass of 
 100 GeV$\lesssim m_{\rm DM} \lesssim$ 1 TeV.
Since the gauge-Higgs dark matter particle can scatter off a nucleon 
 through processes mediated by the Higgs boson 
 in the $t$-channel, a parameter region of our model 
 is constrained by this current experimental bound. 

The elastic scattering cross section between 
 the dark matter and nucleon mediated by the Higgs boson is given as 
\begin{equation}
 \sigma_{el}(DM+N\to DM+N)=\frac{y_{\rm DM}^2 m_N^2 m_{\rm DM}^2}
                          {\pi v_h^2 m_h^4 (m_{\rm DM}+m_N)^2}
                          \left|f_N \right|^2,
\end{equation}
 where $m_N=0.931{\rm eV}$ is the nucleon mass~\cite{PDG}, 
 and $v_h=246{\rm GeV}$ is the VEV of the Higgs doublet. 
The parameter $f_N$ is defined as  
\begin{eqnarray} 
 f_N = \langle N \left| \sigma_q m_q \bar{q}q 
     - \frac{\alpha_s}{4 \pi} G_{\mu \nu} G^{\mu \nu}  
       \right| N \rangle  
     = m_N \left(\frac{2}{9} f_{T_G}+f_{T_u}+f_{T_d}+f_{T_s}\right),
\end{eqnarray} 
where $q$ represents light quarks ($u$, $d$, and $s$) 
 and $G_{\mu \nu}$ is the gluon field strength. 
Contributions from the light quarks to the hadron matrix 
 element is evaluated by lattice QCD simulations~\cite{fbyLattice}, 
\begin{eqnarray}
 f_{T_u}+f_{T_d} \simeq 0.056, \; \; f_{T_s} < 0.038, 
\end{eqnarray} 
while the contribution by gluon $f_{T_G}$ is determined 
 from  the trace anomaly condition~\cite{TrAnom}:
\begin{equation}
  f_{T_G}+f_{T_u}+f_{T_d}+f_{T_s}=1. 
\end{equation}
In our analysis, we use the conservative value, $f_{T_s}=0$. 

For various values of parameters, $\theta_W$ and $c$, 
 with $m_h=120$ GeV fixed, 
 we evaluate the elastic scattering cross sections 
 between the dark matter particle and nucleon. 
\begin{figure}[t]
\begin{center}
\includegraphics[width=0.5\textwidth]{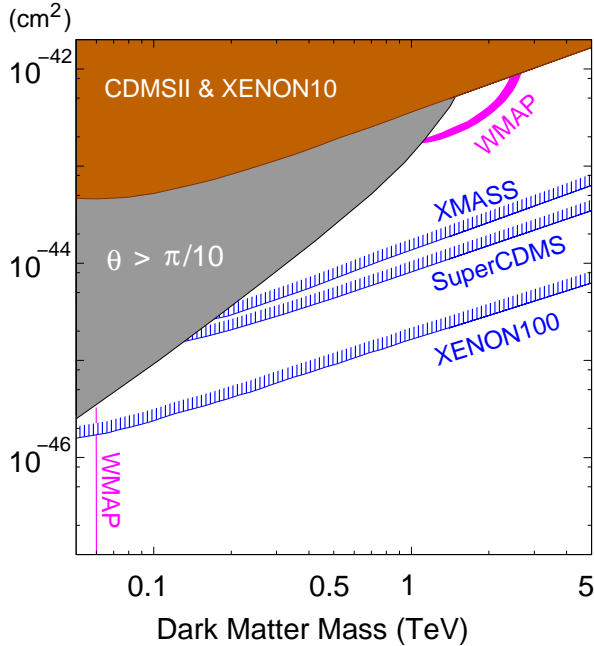}
\end{center}
\caption{{\it The direct detection}:
 The red regions correspond to parameter sets that predict the right abundance. 
 The parameter sets with $\theta_W>\pi/10$  are indicated in gray, and those 
  excluded by the current bound from the direct detection experiments 
  are in brown.
 The expected search limits by future experiments are also shown.}
\label{fig:DD}
\end{figure}
The result is shown in figure \ref{fig:DD}. 
The parameter sets in red regions lead to the appropriate
 dark matter abundances. 
The gray region corresponds to $\theta_W>\pi/10$, which we do not consider 
 as discussed in section~\ref{Sec:constraints}.
The already excluded region 
 from XENON10~\cite{XENON10} and CDMS II~\cite{CDMS} experiments 
 is shown in brown, 
 by which a part of the red region with $m_{\rm DM} =2-3$ TeV 
 is excluded. 
Here, we naively extrapolate the exclusion limit beyond 1 TeV, 
 although the experimental bounds shown in the original papers
 are depicted in the range $m_{\rm DM} \leq 1$ TeV~\footnote{
We would like to thank Yoshitaka Itow for his advise on the 
 current experimental bounds for the dark matter mass beyond 1 TeV.
}. 
The other three lines indicate expected future limits 
 by XMASS~\cite{XMASS}, SCDMS~\cite{SCDMS} and , XENON100~\cite{XENON100}
 respectively from above to below. 
The allowed region with the dark matter mass around TeV is fully 
 covered by the future experiments. 
On the other hand, most of the narrow region consistent 
 with the observed dark matter abundance  
 is out side of the reach of the future experiments. 

\section{Summary} 

In extradimensional theories, 
 the AP boundary condition for a bulk fermion 
 can be imposed in general. 
We show that the lightest mode of the AP fields 
 can be stable and hence become a candidate 
 for the dark matter in the effective 4D theory
 due to the remaining accidental discrete symmetry. 
This mechanism works even with general non-flat metric, 
 in contrast to the KK parity which does not work 
 in a simple warped model.

Although we can introduce the AP fields in various phenomenological 
 extradimensional models, they are usually not so strongly motivated 
 except for providing the dark matter particle. 
In contrast, it is worth noting that in the 
 gauge-Higgs unification scenario, AP fields often play 
 a crucial role to realize a phenomenologically viable model. 
Thus, we examine the possibility of the dark matter 
 in the gauge-Higgs unification scenario. 
We find that due to the structure of the gauge-Higgs unification, 
 the interactions of the dark matter particle 
 with the SM particles, especially with the Higgs boson, 
 are largely controlled by the gauge symmetry
 and the model has a strong predictive power 
 for the dark matter phenomenology. 
Because of this feature, we call this scenario
 as the gauge-Higgs dark matter scenario.

We have investigated this scenario based on 
 a five-dimensional SO(5)$\times$U(1)$_X$ 
 gauge-Higgs unification model compactified 
 on the warped metric as an example. 
This model is favorable because it contains the bulk 
 custodial symmetry and thus a few TeV KK scale 
 can be consistent with the electroweak precision measurement. 
We have evaluated the relic abundance of the dark matter particle 
 and identified the parameter region of the model 
 to be consistent with the observed dark matter relic density. 
We have found two allowed regions: 
One is a quite narrow region where the right dark matter relic 
 density is achieved by the dark matter annihilation 
 through the Higgs boson resonance, 
 so that the dark matter mass is close to a half 
 of the Higgs boson mass. 
In the other region, the dark matter annihilation process 
 is efficient and the dark matter particle with a few TeV mass 
 is consistent with the observations. 
Furthermore, we have calculated the cross section 
 of the elastic scattering between the dark matter particle 
 and nucleon and shown the implication of the gauge-Higgs 
 dark matter scenario for the current and future direct 
 dark matter detection experiments. 
It turns out that the region with a few TeV dark matter mass 
 is partly excluded by the current experiments 
 and the whole region can be explored by future experiments. 
On the other hand, most of the narrow region is out side of the
 experimental reach.

\vspace{1cm}
\leftline{\bf Acknowledgments}

This work is supported in part by a Grant-in-Aid for Science Research from the 
 Ministry of Education, Culture, Sports, Science and Technology, Japan 
 (Nos.~16540258 and 17740146 for N.~H., No.~21740174 for S.~M. and 
  No.~18740170 for N.~O.), 
 and by the Japan Society for the Promotion of Science (T.Y.).

\end{document}